\documentstyle[aps,preprint]{revtex}              
\draft
\title{Gauge invariant nonlinear electric transport in mesoscopic conductors}
\author{T. Christen$^{1)}$ and M. B\"uttiker$^{1),2)}$}
\address{1) D\'epartement de physique th\'eorique,
Universit\'e de Gen\`eve, 24 Quai Ernest-Ansermet \\ 
CH-1211 Gen\`eve, Switzerland\\
2) Institute of Theoretical Physics, University of California\\
Santa Barbara, Ca 93106-4030 USA}
\begin{document}
\maketitle
\newpage
\begin{abstract}
We use the scattering approach to 
investigate the nonlinear
current-voltage characteristic of mesoscopic conductors.
We discuss the leading nonlinearity by taking into account the
self-consistent nonequilibrium potential. We emphasize conservation
of the overall charge and current which are connected to the 
invariance under a global voltage shift (gauge
invariance). As examples, we discuss the rectification coefficient of a
quantum point contact and the nonlinear current-voltage
characteristic of a
resonant level in a double barrier structure.
\end{abstract}
{\em Introduction -} The scattering approach is widely used
to describe electric transport of phase-coherent
mesoscopic conductors in the linear response regime \cite{REV}.
Nonlinear effects of interest are 
asymmetric  
current-voltage characteristics and rectification \cite{TG}, the evolution
of half-integer conductance plateaus \cite{GK}, the breakdown of
conductance quantization \cite{KH}, and negative differential
conductance and hysteresis \cite{KLUK}. In general, in the nonlinear
regime inelastic processes which destroy the phase coherence can play
an important role. Below, however, we emphasize phase-coherent
nonlinear transport. The nonequilibrium state is determined by 
the potential generated by the nonequilibrium charges piled up in
a biased conductor. A reasonable theory of nonlinear
electric transport has to take into account this potential self-consistently.
Most importantly, for a mesoscopic sample there exists a Gauss volume 
which encloses the mesoscopic conductor and nearby gates (capacitors)
such that the electric flux through the surface of this volume 
vanishes \cite{BU1}.
Thus the total charge inside this volume is conserved.
Consequently, 
the I-V-characteristic 
is gauge invariant: it is invariant under 
a global potential shift and depends only on the differences of the voltages
applied to the contacts of the sample or nearby capacitors \cite{BU1}.
Nonlinearities in  
tunnel contacts have already been addressed in an early paper
by Frenkel \cite{FR}. The necessity to consider the self-consistent
potential has been emphasized by
Landauer \cite{LA1}. Nevertheless, many works on nonlinear transport\cite{GK,AK} consider only the
non-interacting case or consider interactions which are not gauge 
invariant.
An exception is, e.g., a work by Kluksdahl
et al. \cite{KLUK}
which provides a self-consistent numeric treatment of a tunneling
barrier with the help of Wigner functions. \\ \indent
In this Letter we present a self-consistent,
gauge-invariant theory of
weakly nonlinear transport using the scattering approach.
A gauge invariant theory of linear ac-conductance is the subject 
of Refs. \cite{BU1,BC1,BTP,CB1}. The discussion of non-linear transport
presented here is based on similar concepts.
We can illustrate the main points using an effective Hartree
approach to treat the effects of the long range Coulomb interaction.
The Hartree approach has a wide range of applicability but 
excludes single-charge effects \cite{CS}. \\ \indent
{\em The leading nonlinearity -}
Consider a mesoscopic conductor which is connected via contacts
$\alpha = 1,...,N$ to $N$ electron reservoirs. We allow that some
parts of the conductor are disconnected from other parts in order
to include into the formalism the presence of nearby gates
(capacitors). The transport
properties are described by the scattering-matrix elements
$s_{\alpha \beta n m } $ which relate the out-going
current amplitude in channel $n$ at contact $\alpha $ to the incident
current amplitude in channel $m$ at contact $\beta $.
We denote by ${\bf s}_{\alpha \beta}$ the scattering matrix
with rows and lines associated with the channels $n$ and $m$ \cite{BU2}. 
The scattering matrix is a function of the energy $E$ of the wave and is
a functional of the electric potential $U(x, \{ V_{\gamma}\})$ in
the conductor. This potential in turn depends on the shifts
$eV_{\gamma}$ of the
electrochemical potentials $\mu_{\gamma}$ of the reservoirs
away from the equilibrium state associated with
$\mu ^{eq}$. Hence, the determination
of the scattering matrix ${\bf s}_{\alpha \beta}(E, \{ V_{\gamma}\})$ 
as a function of the energy and the voltage shifts
in the reservoirs defines
a formidable self-consistent problem. Once this
scattering matrix is found we can calculate the
current through contact $\alpha $ \cite{BU2}
\begin{equation}
I_{\alpha}= \frac{2e}{h}
\sum _{\beta = 1} ^ {N}
\int dE\: f(E-E_{F}-eV_{\beta}) \:
 A_{\alpha \beta}(E,\{ V_{\gamma}\})
\label{current}
\end{equation}
where $f(z)=(1+\exp (z/k_{B}{\rm T}))^{-1}$ is the Fermi function of a
reservoir at temperature T, and $
A_{\alpha \beta}(E,\{ V_{\gamma}\})=  {\rm \bf Tr}
[{\bf 1}_{\alpha} \delta _{\alpha \beta}-
{\bf s}^{\dagger}_{\alpha \beta}(E,\{ V_{\gamma}\}){\bf s}_{\alpha \beta}
(E,\{ V_{\gamma}\})]$ are the {\em screened} (negative) transmission
functions. To discuss weakly nonlinear transport,
we write for the current 
\begin{equation}
I_{\alpha} = \sum _{\beta }G_{\alpha \beta}V_{\beta }+
\sum _{\beta \gamma }G_{\alpha \beta \gamma}V_{\beta}V_{\gamma} + ...\;\;.
\label{secondorder}  
\end{equation} 
The coefficients $G_{\alpha \beta} $ and
$G_{\alpha \beta \gamma}$ are obtained from an expansion
of Eq. (\ref{current}) with respect to the voltages $V_{\alpha}$. 
One obtains for the linear conductance the well-known expression  
$ G_{\alpha \beta}= (2e^{2}/h) \int dE\:(-\partial _{E}f) 
A_{\alpha \beta} $, where the $A_{\alpha \beta} $
are taken at $V_{1}=...=V_{N}=0$.
An expansion of Eq. (\ref{current})
up to ${\cal O}(V_{\beta}V_{\gamma})$ yields 
\begin{equation}
G_{\alpha \beta \gamma}= \frac{e^{2}}{h}
\int dE\: (-\partial _{E}f)
\left( 2\partial _{V_{\gamma}}A_{\alpha \beta} +
e \partial _{E}A_{\alpha \beta}  \delta_{\beta \gamma}
\right)\;\;.
\label{Gabc1}
\end{equation}
Conservation of the total charge (gauge invariance)
implies that the currents are conserved \cite{BU1,BC1}
and that the currents be independent of
a global voltage shift $W$: $V_{\alpha}\to V_{\alpha}+W $.
These conditions imply the sum rules \cite{BTP}
$\sum _{\alpha} G_{\alpha \beta} =$
$\sum _{\beta} G_{\alpha \beta} =0$. Similarly, the
coefficients of the second-order nonlinearity must obey
\begin{equation}
\sum _{\alpha} G_{\alpha \beta \gamma}=
\sum _{\gamma} (G_{\alpha \beta \gamma}+
G_{\alpha \gamma \beta })=0 \;\;.
\label{sumrules}
\end{equation} 
The sum rules also follow
from the unitarity of the scattering matrix, i.e.
$\sum _{\alpha} A_{\alpha \beta} =$ $\sum _{\beta} A_{\alpha \beta} =
0$, and from the gauge invariance condition
$e\partial _{E} A_{\alpha \beta} +\sum _{\gamma}
\partial _{V_{\gamma}}A_{\alpha \beta} =0$.\\ \indent
To determine the derivatives $\partial _{V_{\gamma}} A_{\alpha
\beta} $ we recall that the $A_{\alpha \beta}$ are
functionals of the electric potential $U(x,\{V_{\gamma}\})$. 
Within linear response theory the potential variation
$\Delta U(x)= U(x)-U_{eq}(x)$
away from equilibrium is determined by the {\em characteristic
potentials} $u_{\gamma}(x)=
(\partial U(x)/\partial V_{\gamma})_{eq}$. 
To derive the characteristic potentials we introduce the
{\em injectivities} $dn(x,\alpha)/dE=$ $ -(1/4\pi i)
\sum _{\beta} [ {\bf s}_{\beta \alpha }^{\dagger}(\delta {\bf s}_{\beta \alpha 
}/e\delta U(x)) -
(\delta {\bf s}_{\beta \alpha
}^{\dagger}/e\delta U(x)) {\bf s}_{\beta \alpha }] $.
The injectivity which is related to a dwell-time \cite{BC1,GCB}
is a partial local density of states,
i.e. the sum over all contacts $\alpha $ gives the local density of
states $dn(x)/dE$. Variations $eV_{\alpha}$ of the electrochemical
potentials $\mu _{\alpha}$ in the contacts at fixed electric
potential $U_{eq}(x)$ in the conductor induce a charge density 
$q^{(1)}(x) = e^{2} \sum _{\alpha} (dn(x,\alpha)/dE)\: V_{\alpha} $.
In general, however, the electric potential changes by
$\Delta U(x)$ which induces an additional charge density $q^{(2)}(x) = -
e^{2}\int \Pi (x,y) \Delta U(y) d^{3}y $,
where $\Pi (x,y)$ is the Lindhard polarization function.
Inserting the total nonequilibrium charge
$q^{(1)}+q^{(2)}$ and the potential variation $\Delta U =\sum
u_{\alpha} V_{\alpha}$ into the Poisson equation
yields for the characteristic potentials an equation with a nonlocal
screening term \cite{BU1} 
\begin{equation}
-{\em \Delta } u_{\alpha }(x)+ 4\pi e^{2} \int d^{3}y \: \Pi (x,y) u_{\alpha
}(y)=
4\pi e^{2} \frac{dn(x, \alpha)}{dE} \;\;.
\label{poisson}
\end{equation}
For simplicity, the dielectric constant is to be unity.
The derivative $ \partial _{V_{\gamma}}A_{\alpha \beta}$
can be expressed in terms of the functional derivative 
$\delta A_{\alpha \beta}/\delta U(x) $ and the characteristic
potentials, which yields for Eq. (\ref{Gabc1}) 
\begin{equation}
G_{\alpha \beta \gamma}= \frac{e^{3}}{h} \int dE\: (-\partial _{E}f)
\int \: d^{3}x\: \frac{\delta A_{\alpha \beta}}{e\delta U(x)}
\left( 2u_{\gamma}(x) - \delta_{\beta \gamma}\right)\;\;.
\label{Gabc2}
\end{equation}
{\em Discrete potential model -}
We can obtain analytical results
by using a convenient coarse graining
model for the conductors \cite{CB1}. We discretize the conductor
by decomposing it into $M$ regions where the electrostatic potential
and the charge density are assumed to be constant. Space
dependent quantities are written as $M$-dimensional
vectors or as $M\times M$-matrices. For example, the electrostatic
potential variation $\Delta U(x)$ becomes a vector
$\Delta {\bf  U} =(\Delta U_{1},...,\Delta U_{M})$, where $\Delta U_{k}$
denotes the electrostatic potential shift in region $k$.
We write the injectivities as vectors ${\bf D}_{\alpha}=
e^{2}(dN_{1 \alpha}/dE,...,dN_{M \alpha}/dE)$ where we multiplied
by $e^{2}$ for later convenience. The functional derivative
with respect to the potential $U(x)$ is replaced by
the gradient ${\bf \nabla_{U}} $, and
the polarization function is a matrix ${\bf \Pi}$.
We introduce a geometric capacitance matrix ${\bf C}$ associated with these
regions and which is calculated from the Poisson equation. 
The nonequilibrium charge distribution is ${\bf q}=
{\bf C}\Delta  {\bf U} =
\sum_{\beta }{\bf D}_{\beta}V_{\beta}-{\bf \Pi}\Delta { \bf U}$ which
yields for the characteristic potentials ${\bf u}_{\beta}= ({\bf
\Pi}+{\bf C})^{-1}{\bf D}_{\beta}$. Hence, Eq. (\ref{Gabc2}) becomes
\begin{equation}
G_{\alpha \beta \gamma}= \frac{e^{2}}{h}\int dE (-\partial _{E}f)
\bigl( 2{\bf \nabla _{U}}
A_{\alpha \beta}\: ({\bf \Pi}
+ {\bf C})^{-1}{\bf D}_{\gamma}  -
e \delta _{\beta \gamma} \partial _{E} A_{\alpha \beta} \bigr)\;\;.
\label{Gabcdiscrete}
\end{equation}
We mention that the charge $q_{k}$ in region $k$ can be written
in terms of an electrochemical capacitance matrix \cite{CB1},
$q_{k}=\sum _{\beta} C_{k\beta }^{\mu}V_{\beta}$ where
$C_{k\beta }^{\mu}= \sum _{l} C_{kl}u_{l\beta}$.\\ \indent
In the following we investigate two-terminal devices.
Due to gauge invariance, the current
$I_{1}=-I_{2}= G_{11}\Delta V + G_{111}(\Delta V)^{2} + ...  $
depends only on the voltage
difference $\Delta V= V_{1}-V_{2}$. The second-order nonlinearity is
thus determined by the single
rectification coefficient $G_{111}$.\\ \indent
{\em Rectification of a quantum point contact -} 
We follow closely Ref. \cite{CB1} and discretize the quantum point
contact (QPC) by defining to the left and to the right of the constriction
two regions $\Omega _{1}$ and $\Omega _{2}$
with sizes of the order of the screening length. We assume zero
temperature and take all quantities at the Fermi energy.
A voltage difference across the QPC induces a dipole
consisting of charges $q_{1}$ and $q_{2}=-q_{1}$ which
reside in $\Omega _{1}$ and $\Omega _{2}$, respectively.
Each occupied subband $j$ contributes
with a transmission probability $T^{j}$ to the total transmission
function ${\cal T}=\sum _{j} T^{j}$. Due to charge conservation,
the geometrical capacitance matrix has the specific form
$C_{kl}=(-1)^{k+l} C $. 
Assuming a barrier potential of the constriction which allows a WKB
approximation, we can neglect nonlocalities and write
$\Pi _{kl}=D_{k}\delta _{kl}$. With each channel $j$ one can
associate a density of states $D_{k}^{j}$ in region $\Omega _{k}$,
and it must hold $D_{k}=\sum _{j}D_{k}^{j}= \sum _{\alpha} D_{k\alpha }$. 
For a voltage variation $V_{1}$, channel $j$ injects from
contact $1$ into $\Omega _{2}$ a charge $D_{2}^{j}(T^{j}/2) V_{1}$
and into $\Omega _{1}$ a charge $D_{1}^{j}(1-T^{j}/2)V_{1}$.
A sum over all channels $j$ yields the injectivity ${\bf D}_{1}$
of contact $1$ with the components $D_{11}= D_{1}(1-T_{1}/2)$ and
$D_{21}=D_{2}T_{2}/2$.
Here, $T_{k}=D_{k}^{-1}\sum _{j}D_{k}^{j}T^{j}$ are
averaged transmission probabilities.
We obtain for the characteristic potentials 
$u_{11}= 1-T_{1}/2-C^{\mu}/D_{1}$ and $u_{21}=
T_{2}/2+C^{\mu}/D_{2}$, where the electrochemical capacitance
is given by $C^{\mu}=R/(C^{-1}+D_{1}^{-1}+D_{2}^{-1})$
and where we introduced the average reflection probability
$R=1-(T_{1}+T_{2})/2$. The rectification coefficient becomes thus
\begin{equation}
G_{111}= \frac{e^{2}}{h}\:\left(
\frac{C^{\mu}}{C} (\partial _{U_{1}} {\cal T}- \partial _{U_{2}}{\cal T})
+[C^{\mu}(D_{2}^{-1} -D_{1}^{-1})-\frac{1}{2}(T_{1}-
T_{2})] e \partial _{E} {\cal T} \right)\:\:.
\label{Gabcqpc}
\end{equation}
As one expects, $G_{111}$ vanishes for a symmetric barrier.
Asymmetries can be due to differences of $D_{k}$, $T_{k}$,
and $\partial _{U_{k}}{\cal T}$ for different $k$.
Consider a single-channel conductor where
$T_{1}-T_{2}$ vanishes and which can be seen as a model
for a one-dimensional conductor with an impurity.
In the limit $C\ll D_{k} $, charges are screened and
the rectification coefficient is 
$G_{111} = (e^{2}/h) R(\partial _{U_{1}} {\cal T}-
\partial _{U_{2}}{\cal T})$. On the other hand, in the limit
$D_{k} \ll C$, Eq. (\ref{Gabcqpc}) gives
$G_{111} = (e^{3}/h) (D_{2}-D_{1})R \partial _{E}T
/(D_{1}+D_{2})$ which is proportional to the asymmetry of the
density of states. Clearly, with a discretization by more than two
regions, an additional source of asymmetry can result from asymmetric 
electric screening (unequal capacitance
coefficients).\\ \indent
{\em Resonant tunneling barrier -}
To model the resonant tunneling barrier, we consider three discrete
potential regions denoted by indices $0$, $1$, and $2$ which
correspond to the well, the left side of the barriers,
and the right side of the barriers, respectively.
We assume fully screened charges. Consider a single resonant
level with energy $E_{r}+eU_{0}$ which is clearly
separated from other resonant levels, from the band bottoms, and from
the left and right barrier tops. The linear conductance of such a
structure is given by $G_{11}=
(2e^{2}/h)T$ where $T$ is the transmission probability of the double
barrier structure. Linear response is only valid if the bias
$|e \Delta V|$ is much smaller than the width $\Gamma $ of the
resonance. To find the nonlinear
characteristic, we restrict ourselves to near resonant conditions
where the transmission probability can be approximated by the
Breit-Wigner formula  $T(E,U)= \Gamma _{1} \Gamma _{2}/|\Delta
|^{2}$ with $|\Delta |^{2} = (E-E_{r}-eU_{0})^{2} +\Gamma ^{2}/4$.
In this range the nonlinearity is solely due to the resonance.
In particular, we assume that $\Gamma _{1} $ and $\Gamma _{2}$, and thus
the width $\Gamma =\Gamma _{1} +\Gamma _{2}$ of the resonant level
and the asymmetry, $\Delta \Gamma = \Gamma _{1}-\Gamma _{2}$,
are constant. The injectivities are \cite{BTP} $D_{0 \alpha}(E,U_{0}) =
(e^{2}/2\pi)\Gamma _{\alpha} /|\Delta |^{2}$. The relation between the
electrostatic potential shift $\Delta U_{0}$ and the voltage shifts
$V_{\alpha} = (\mu _{\alpha}-\mu ^{eq})/e $ $(\equiv \Delta U_{\alpha}) $ 
is obtained from the charge-neutrality condition 
\begin{equation}
\int _{-\infty}^{\mu_{1}} D_{01}(E,U_{0})\: dE+
\int _{-\infty}^{\mu_{2}} D_{02}(E,U_{0}) \: dE
-\int _{-\infty}^{\mu ^{eq}} D_{0}(E,U_{0}^{eq}) \: dE \equiv 0
\;\;,
\label{cneutr}
\end{equation}
which can be integrated analytically. Eq. (\ref{cneutr})
determines $W=\Delta U_{0}-(V_{1}+V_{2})/2$ which depends on 
$\Delta V$ only. Equation (\ref{current}) yields 
for the current $I\equiv I_{1}=-I_{2}$
\begin{equation}
I=\frac{4e}{h} \frac{\Gamma _{1}\Gamma _{2}}{\Gamma}
\left( {\rm arctan}[2
(\Delta E - e (W-\Delta V/2))/\Gamma ] -
{\rm arctan}[2 (\Delta E - e (W+\Delta V/2))/\Gamma ]   \right) \;\;,
\label{IVRTB}
\end{equation} 
where $\Delta E = \mu _{eq}-(eU_{0}^{eq}+E_{r})$ is the equilibrium
distance between the Fermi energy and the resonance.
Without considering the self-consistent shift $\Delta U_{0}$ 
one would get a wrong result which is not gauge invariant.
The current given by Eq. (\ref{IVRTB}) saturates at a maximum value
proportional to $\pi /2 - {\rm arctan}(2\Delta E /\Gamma)$.
The conduction is optimal for $\Delta E =0$ and
$\Gamma _{1}=\Gamma _{2}$ when $I= 2(e/h)\Gamma {\rm arctan }(e\Delta
V/\Gamma)$. In Fig. \ref{fig1} we have plotted the characteristic
as a function for an asymmetry $\Delta \Gamma /\Gamma = -1/3$ and
for various values of $\Delta E$.  
For the case of complete screening considered here,
the resonant level floates up or down
to keep the charge in the well constant. An expansion of the current
yields $G_{111}=- (e^{3}/h)(\Delta \Gamma /\Gamma)\partial _{E}T$
which is in accordance with (\ref{Gabcdiscrete}) and with 
Ref. \cite{CB1} (thin dotted lines in Fig. \ref{fig1}). 
The case of incomplete screening can similarly be treated with 
our approach. In this case, at large voltages, the 
resonance can eventually fall below the conductance band bottom
of the injecting reservoir as is known from the semiconductor
double barrier structures. In the general case, even an elastically
symmetric resonance can be rectifying if the electrical screening
is asymmetric.\\ \indent
{\em Conclusion -} In this work we have emphasized that a reasonable
theory of nonlinear electrical transport in mesoscopic conductors
requires a self-consistent
treatment of the long-range Coulomb interactions which insures gauge
invariant results. We showed that the
scattering approach to mesoscopic conduction provides an appropriate
method if one introduces screened scattering matrices which depend
on the self-consistent potential.\\ \indent 
{\em Acknowledgement}
This work has been supported by the Swiss National Science
Foundation. For M.B. part of this work was supported by the
US National Science Foundation under grant PHY94-07194. 

\begin{figure}
\caption{Current-voltage characteristic of a single level
of the resonant tunneling barrier; $\Gamma _{1}=0.5 meV $,
 $\Gamma _{2}= 1meV$, $\Delta E= 0$ (solid),
$\Delta E= -\Gamma$ (dashed); $\Delta E= -2\Gamma $ (dotted).
Thin dotted lines indicate the quadratic approximation.}
\label{fig1}
\end{figure}

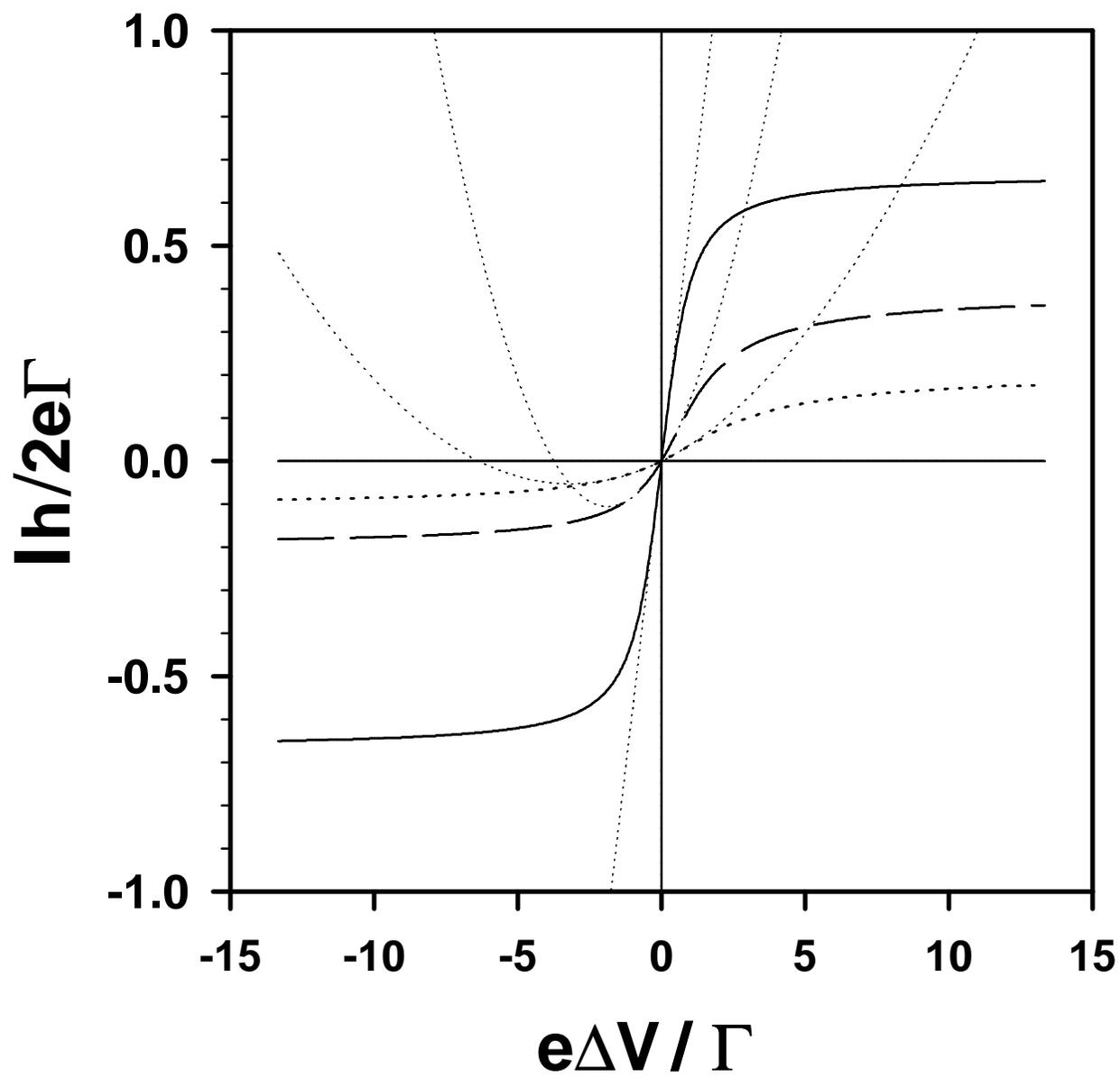
\end{document}